\documentclass[aps,pre,twocolumn,titlepage]{revtex4}

\usepackage{graphicx}% Include figure files
\usepackage{color}
\usepackage{dcolumn}% Align table columns on decimal point
\usepackage{latexsym}
\usepackage{hyperref,amssymb}
\usepackage{url}
\usepackage{color,soul}
\usepackage{graphicx}
\usepackage{verbatim}
\usepackage{multirow}
\usepackage{amsmath}
\newcommand{\beq}{\begin{eqnarray}}
\newcommand{\eeq}{\end{eqnarray}}
\usepackage{mathrsfs}
\usepackage{float,soul}
\usepackage[dvipsnames]{xcolor}
\usepackage{slashed}
\usepackage{graphicx}   % need for Fig.ures
\usepackage{epstopdf}
\usepackage{hyperref}   % use for hypertext links, including those to external documents and URLs
\usepackage{hyperref}
\hypersetup{colorlinks,
}

\usepackage[latin1]{inputenc}

\begin{document}

\title{Observation of fast sound in two-dimensional dusty plasma liquids}

	\author{Zhenyu Ge$^{1}$}
	\author{Dong Huang$^{1}$}
	\author{Shaoyu Lu$^{1}$}
	\author{Chen Liang$^{1}$}
	\author{Matteo Baggioli$^{2}$}
	\thanks{b.matteo@sjtu.edu.cn}
	\author{Yan Feng$^{1}$}
	\thanks{fengyan@suda.edu.cn}
	\affiliation{$^{1}$ Institute of Plasma Physics and Technology, School of Physical
Science and Technology, Jiangsu Key Laboratory of Thin Films, Soochow University, Suzhou 215006, China\\
$^{2}$ Wilczek Quantum Center, School of Physics and Astronomy, Shanghai Jiao Tong University, Shanghai 200240, China \& Shanghai Research Center for Quantum Sciences, Shanghai 201315, China\\
	}

\date{\today}

\begin{abstract}
Equilibrium molecular dynamics simulations are performed to study two-dimensional (2D) dusty plasma liquids. Based on the stochastic thermal motion of simulated particles, the longitudinal and transverse phonon spectra are calculated, and used to determine the corresponding dispersion relations. From there, the longitudinal and transverse sound speeds of 2D dusty plasma liquids are obtained. It is discovered that, for wavenumbers beyond the hydrodynamic regime, the longitudinal sound speed of a 2D dusty plasma liquid exceeds its adiabatic value, \emph{i.e.}, the so-called \textit{fast sound}. This phenomenon appears at roughly the same length scale of the cutoff wavenumber for transverse waves, confirming its relation to the emergent solidity of liquids in the non-hydrodynamic regime. Using the thermodynamic and transport coefficients extracted from the previous studies, and relying on the Frenkel theory, the ratio of the longitudinal to the adiabatic sound speeds is derived analytically, providing the optimal conditions for fast sound, which are in quantitative agreement with the current simulation results. 

\end{abstract}

\maketitle

\section{Introduction}
Hydrodynamics \cite{Landau:2013,Boon:1991} is an effective description of the low-energy collective dynamics of a certain physical system, whose application is very general and indeed not restricted to fluids \cite{Martin:1972,Baggioli:2023,Halperin:1969}. The regime of validity of such a framework is the so-called \textit{hydrodynamic limit}, which corresponds to long length-scale and time-scale processes, i.e., the macroscopic coarse-grained ``world''. The low-energy degrees of freedom, which constitute the building blocks for hydrodynamics, are provided by the collective hydrodynamic modes, whose number can be uniquely determined by looking at the symmetries of the system (\emph{i.e.}, number of conserved quantities and Goldstone modes).

The dispersion relation of the hydrodynamic modes can be extracted using time correlation functions or directly the Fourier transformed dynamic structure factor $S(k,\omega)$ (where $k,\omega$ are respectively the wavenumber and the frequency), which can be measured using inelastic X-ray scattering (IXS) or inelastic neutron scattering (INS) experiments. In the hydrodynamic limit, classical liquids exhibit only a pair of propagating longitudinal modes, usually denoted as (longitudinal) sound, with dispersion relation of
\begin{equation}
    \omega=\pm v_s k -i\frac{\Gamma_L}{2}k^2 +\dots
\end{equation}
where the ellipsis indicates higher-order terms in the wavenumber $k$. Here, $v_s$ is the adiabatic speed of sound, and $\Gamma_L$ the sound attenuation constant, whose expressions can be derived within the hydrodynamic theory. Differently from solids \cite{Chaikin:2000}, liquids do not display propagating shear waves in the hydrodynamic limit, as a consequence of a vanishing static shear modulus.

Beyond the hydrodynamic limit, liquids display a complex dynamics and several interesting features which are still debate topics. As a matter of fact, it remains unclear whether the definition itself of liquids, in contraposition to solids, is meaningful at all beyond the hydrodynamic regime. First, above a certain cutoff wavenumber, sometimes indicated as the \textit{k-gap}, propagating shear waves are observed in liquids \cite{Trachenko:2015,Baggioli:2020}. The detection of propagating non-hydrodynamic shear waves has been directly reported in many simulations (\emph{e.g.}, \cite{Yang:2017}), and even in experiments (\emph{e.g.}, \cite{Nosenko:2006}). This feature is often used to argue for the solid-like nature of liquids beyond the hydrodynamic limit. In jargon, for fast enough processes or short enough length-scales, liquids are qualitatively indistinguishable from solids \cite{Noirez:2012}. Second, the spectrum of instantaneous vibrational modes in liquids displays the presence of unstable (instantaneous) normal modes (INMs), with purely imaginary frequencies \cite{Keyes:1997,Stratt:1995}. INMs might play a fundamental role in self-diffusion \cite{Xiong:1999}, the low-frequency behavior of the liquid density of states \cite{Zaccone:2021,Stamper:2022}, and possibly also thermodynamic properties such as the heat capacity \cite{Madan:1993,Baggioli:2021}.

Finally, and most importantly for this work, the longitudinal sound mode in liquids displays the so-called \textit{fast sound} phenomenon for wavnumbers beyond the hydrodynamic limit. In particular, the longitudinal sound speed exceeds the hydrodynamic value, which is given by the adiabatic sound speed of
\begin{equation}\label{adi}
    v_s^2=\frac{K_s}{\rho}\,,
\end{equation}
where $K_s$ is the adiabatic bulk modulus and $\rho$ the mass density \cite{Hansen:2013}. Interestingly, the onset of fast sound propagation coincides roughly with the appearance of propagating shear waves, and it is therefore often thought as a manifestation of the emergent solidity of liquids beyond the hydrodynamic limit. Supporting this statement, the speed of sound in the fast sound regime is approximately equal to the expected expression in solids which is given by
\begin{equation}\label{lala}
    {v'_L}^2=\frac{K_s+\mu_\infty}{\rho}\,,
\end{equation}
where $\mu_\infty$ is the instantaneous shear modulus, not to be confused with the static shear modulus which vanishes in liquids.

In liquid water, the existence of fast sound was first observed in MD simulations by Rahman and Stillinger in 1974 \cite{Rahman:1974}, and later observed experimentally using INS by Teixeira and collaborators \cite{Teixeira:1985}, and using IXS by Sette and collaborators \cite{Sette:1995}. These results sparkled a huge amount of simulation, theoretical and experimental effort which is nicely summarized for the case of water in \cite{Ruocco:2008} (see also \cite{Balucani:1993,Cunsolo:1999,Cunsolo:2015}). Nowadays, fast sound (also labelled \textit{positive sound dispersion}, PSD) has been observed and discussed in a plethora of liquids (\emph{e.g.}, \cite{Balucani:1993,Bryk:2005,Bosse:1986,Demmel:2004,Alvarez:1998,Campa:1988,Ruzicka:2004,Bencivenga:2014,Sabirov:2010,Bryk:2010,Santucci:2006,Bryk:2014}).

In order to explain these phenomena, the classical hydrodynamic framework is of no help and it has to be augmented. Several generalization schemes have been proposed in the past, including the generalized collective mode theory \cite{Bryk:1997,Bryk:2002,Bryk:2011}, the memory function methods \cite{Cohen:1971}, the generalized hydrodynamic theory \cite{Schepper:1988,Mryglod:1999}, the solid-like theories of liquids \cite{Trachenko:2015,Baggioli:2020}, the topological structures \cite{Baggioli:2022}, and many more. A common denominator to all these approaches is the necessity to introduce additional non-hydrodynamic degrees of freedom, whose frequency $\omega$ does not vanish upon sending the wavenumber to zero, $k \rightarrow 0$. Since the purpose of this work is mostly observational, we will not indulge at length in the details of any of these theories. On the contrary, we will limit ourselves to interpret our observations with the simplest theoretical framework, the Frenkel theory \cite{Frenkel:1955}, whose details will be briefly explained in the following sections.

Dusty plasma~\cite{Thomas:1996,Lin:1996,Melzer:1996,Merlino:2004,Kalman:2004,Morfill:2009,Bonitz:2010}, also known as complex plasma, typically refers to partially ionized gas containing micron-sized dust particles, which are highly charged to $\sim -10^4 e$. Laboratory dusty plasma is an excellent model system that provides a powerful diagnostic of the tracking of individual dust particles. In the laboratory conditions, these highly charged dust particles can be levitated by the electric field in the plasma sheath, and then self-organize into a single layer suspension~\cite{Thomas:2005,Feng:2010,Feng:2012}, \emph{i.e.}, a two-dimensional (2D) dusty plasma. A Yukawa repulsive potential~\cite{Konopka:2000} can be used to describe the interaction between dust particles in this 2D plane. Due to their large charges, the potential energy between dust particles is much larger than their averaged kinetic energy, so that they are strongly coupled~\cite{Merlino:2004,Kalman:2004,Bonitz:2010,Morfill:2009}. As a result, a collection of these dust particles exhibits typical collective behaviors akin to those in liquids~\cite{Feng:2010,Feng:2012} and solids~\cite{Nunomura:2002,Nosenko:2008,Hartmann:2010,Couedel:2010,Kahler:2012,Hartmann:2014,Gogia:2017,Hartmann:2019,ChengRan:2019}. Various fundamental physical phenomena characteristic of solids and liquids are investigated using dusty plasmas at the individual particle level, such as wave propagation~\cite{Nunomura:2003,Goree:2012,Piel:2006,Nunomura:2000,Nunomura:2002_2,Nosenko:2006}, transport properties~\cite{Nunomura:2005,Nosenko:2008,Feng:2012,Feng:2012_2,Juan:1998}, and critical dynamics~\cite{Feng:2008}. 

Wave propagation in dusty plasmas has been intensively investigated using experiments~\cite{Nunomura:2000,Nunomura:2002_2,Nosenko:2006,Nunomura:2002}, simulations~\cite{Ohta:2000,Liu:2003}, and theories~\cite{Wang:2001}. In 2D dusty plasmas, longitudinal~\cite{Nunomura:2002_2} and transverse waves~\cite{Nunomura:2000,Nunomura:2002_2,Piel:2006,Nosenko:2006} can be generated using a lower level disturbance of the laser manipulation, and then their propagation is easily observed at the individual particle level, so that their speeds can be determined~\cite{Nunomura:2000,Nunomura:2002_2,Piel:2006}. If a higher level disturbance is applied, such as an electric pulse~\cite{Samsonov:2004}, or a high speed electric exciter~\cite{Kananovich:2020}, then a shock can be generated in a 2D dusty plasma. For 2D dusty plasma solids or liquids, even without any external disturbance, the stochastic thermal motion of individual dust particles \emph{per se} exhibits wave spectra~\cite{Nunomura:2002}, from which the sound speeds can be determined also, as studied in this paper.

In this work, we investigate the presence of the fast sound phenomenon in 2D dusty plasma liquids using molecular dynamical (MD) Yukawa simulations. In Sec.~\ref{sec2}, we briefly describe the simulation method for 2D dusty plasma liquids, as well as the method to obtain the wave spectra from the stochastic thermal motion of particles. In Sec.~\ref{sec3}, we report our calculated wave spectra and the determined dispersion relations. By comparing the dispersion relations with the adiabatic sound speed of the system, we find the occurrence of the fast sound in 2D dusty plasma liquids. In Sec.~\ref{sec4}, we interpret the discovered fast sound using the Frenkel theory, and then we analytically derive the ratio of the longitudinal to the adiabatic sound speeds, well agreeing with the simulation results. Finally, in Sec.\ref{sec5}, we provide a brief summary.

\section{Methods}\label{sec2}

Traditionally~\cite{Fortov:2005,Morfill:2009}, two dimensionless parameters are often used to characterize 2D dusty plasma systems, which are the coupling parameter $\Gamma = Q^2/(4 \pi \epsilon_0 a k_B T)$ and the screening parameter $\kappa = a / \lambda_D$~\cite{Ohta:2000_2,Sanbonmatsu:2001}. Here, $Q$ is the charge of each particle, $T$ is the average kinetic temperature of particles, $a=1/\sqrt{\pi n}$ is the Wigner-Seitz radius~\cite{Kalman:2004} for the areal number density of particles $n$, and $\lambda_D$ is the Debye screening length. Intuitively, $\Gamma$ is equivalent to the inverse of the effective temperature, while $\kappa$ can be regarded as the length scale of system, since $\lambda_D$ is the environment parameter. Note that, for 2D dusty plasmas, the melting point $\Gamma_m$ varies with the screening parameter $\kappa$~\cite{Hartmann:2005}, while the transition between the liquid-like to gas-like states occurs at $\approx \Gamma_m/20$~\cite{Huang:2023}.

We use MD simulations to obtain the equilibrium dynamics of 2D dusty plasma liquids. The equation of motion~\cite{Huang:2023} for each particle is given by
\begin{equation}\label{MD}
m\ddot{\mathbf{r}}_i=-\nabla\Sigma_j\phi_{ij}\,.
\end{equation}
Here, $\phi_{i j}=Q^{2}\mathrm{exp}(-r_{i j}/\lambda_{D})/4\pi\epsilon_{0}r_{i j}$ is the Yukawa repulsion~\cite{Liu:2003_2} between the particles $i$ and $j$, where $r_{i j}$ is the distance between these two particles.

To improve the statistics of the analysis, our simulation system contains $N=16384$ particles, confined in a $243.8a \times 211.1a$ rectangular box with the periodic boundary conditions. After the simulation system reaches its equilibrium state, we record the particle positions and velocities for the duration of $4.5\times10^6$ steps, with each time step of $0.005 \omega_{pd}^{-1}$, Here, $\omega_{p d}=(Q^2/2\pi\varepsilon_0m a^3)^{1/2}$ is the nominal dusty plasma frequency~\cite{Kalman:2004}. We specify the $\kappa$ values as $1$ and $2.5$ in our simulations, while keeping the same reduced coupling parameter~\cite{Huang:2023} of ${\Gamma}/{\Gamma_m}=0.95$ for both $\kappa$ values. Other simulation details are the same as in~\cite{Feng:2016}.

For our simulated Yukawa liquids, the phonon spectra are obtained using the Fourier transform of the autocorrelation functions for the longitudinal and transverse currents. The longitudinal autocorrelation function is defined as~\cite{Liu:2003,Ohta:2000}
\begin{equation}\label{phononl}
C_L(\boldsymbol{k},t)=\frac{1}{N}\langle[\boldsymbol{k}\cdot\boldsymbol{j}(\boldsymbol{k},t)][\boldsymbol{k}\cdot\boldsymbol{j}(-\boldsymbol{k},0)]\rangle ,
\end{equation}
while the transverse autocorrelation function as
\begin{equation}\label{phonont}
C_T(\boldsymbol{k},t)=\frac{1}{2N}\langle[\boldsymbol{k}\times \boldsymbol{j}(\boldsymbol{k},t)]\cdot[\boldsymbol{k}\times \boldsymbol{j}(-\boldsymbol{k},0)]\rangle .
\end{equation}
Here, $\boldsymbol{k}$ is the wavenumber, while
\begin{equation}
    \boldsymbol{j}(\boldsymbol{k},t)=\sum_{j=1}^N \dot{\mathbf{r}}_j\exp\left[i\boldsymbol{k}\cdot\boldsymbol{r}_j(t)\right]
\end{equation} is the current for a given wavenumber $\boldsymbol{k}$, where $\dot{\mathbf{r}}_j$ and $\boldsymbol{r}_j(t)$ are the velocity and position of the particle $j$, respectively. 
Then we perform the Fourier transform of the autocorrelation functions of Eqs.~(\ref{phononl}) and (\ref{phonont}) to obtain the corresponding phonon spectra as
\begin{equation}\label{phononlt}
\tilde{C}_{L,T}(\boldsymbol{k},\omega)=\int_0^\infty e^{-iwt}C_{L,T}(\boldsymbol{k},t)\:dt\,.
\end{equation}
Here, $\tilde{C}_{L}$ and $\tilde{C}_{T}$ refer to the longitudinal and transverse spectra, respectively.

\section{Results}\label{sec3}

\begin{figure}
\centering
\includegraphics{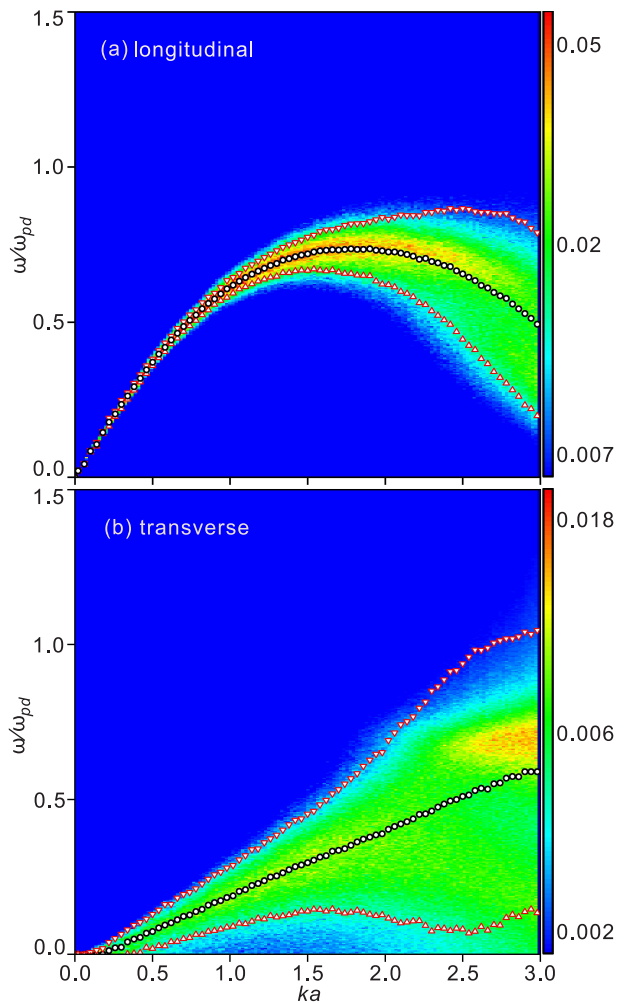}
\caption{\label{fig1} (Color online).
Longitudinal (a) and transverse (b) phonon spectra of the 2D dust plasma liquid for $\kappa=1$ and ${\Gamma}/{\Gamma_m}=0.95$, calculated from equilibrium MD simulations. Circle and triangular symbols on the phonon spectra correspond respectively to ${\omega}_{0}(k)$ and ${\omega}_{0}(k)\pm \gamma_0(k)$ as defined in the DHO model, Eq.\eqref{dho}.}
\end{figure}

The phonon spectra calculated from the stochastic thermal motion of particles~\cite{Nunomura:2002} in the equilibrium MD simulation of 2D dusty plasmas under the conditions of $\kappa=1$ and ${\Gamma}/{\Gamma_m}=0.95$ are plotted in Fig.~\ref{fig1}. The simulation conditions correspond to a liquid state, so that the simulated 2D dusty plasma is isotropic~\cite{Li:2018}, \emph{i.e.}, the phonon spectra corresponding to different directions of the wavenumbers should not differ. Thus, to improve the signal-to-noise ratio (SNR) of the phonon spectra, we take the mean of the phonon spectra by varying the wavenumber direction from $0$ to $\pi/6$ in $16$ steps, since the 2D triangular lattice is symmetric as the system rotates each $\pi/3$ angle around one particle. Then, to further improve the SNR, we average the mean phonon spectra from $10$ independent simulation runs with the same time duration of $t \omega_{pd} = 22,500$.

From our obtained phonon spectra in Fig.~\ref{fig1}, as the wavenumber increases gradually, the corresponding frequency of the transverse spectra increases monotonically. However, the frequency of the longitudinal spectra first increases with the wavenumber to its peak value of $\omega/\omega_{pd} \approx 0.7$, then gradually decreases. In addition, as the wavenumber increases, the width of the frequency distribution becomes wider, for both longitudinal and transverse spectra, as expected from hydrodynamics \cite{Chaikin:2000}. To quantitatively describe the collective dynamics, the dispersion relations of the longitudinal and transverse waves are typically used, and can be determined from the obtained phonon spectra by choosing the peak value of the frequency for each wavenumber, as in~\cite{Ohta:2000}.

To determine dispersion relations, the damped harmonic oscillator (DHO) model is often used to fit the corresponding frequency spectrum, as in~\cite{Yurchenko:2018,Khrapak:2018,Khrapak:2019}. The expression of the DHO model is
\begin{equation}\label{dho}
f(\omega)\propto\frac{1}{\left(\omega-\omega_{0}\right)^{2}+\gamma_{0}^{2}}+\frac{1}{\left(\omega+\omega_{0}\right)^{2}+\gamma_{0}^{2}}\,,
\end{equation}
where the two fitting parameters are the peak frequency $\omega_{0}$ and the width of the frequency distribution $\gamma_{0}$. Due to the noise fluctuation in the obtained phonon spectra, directly choosing the peak frequency may include substantial systematic errors. However, using the DHO model to fit to the obtained spectrum is able to greatly suppress the effect of the random noise fluctuations in the spectrum. As shown in Fig.~\ref{fig1}, the obtained values of the peak frequency $\omega_{0}$ vary smoothly with $k$, while the width of the frequency distribution $\gamma_{0}$ becomes larger monotonically.

\begin{figure}
\centering
\includegraphics{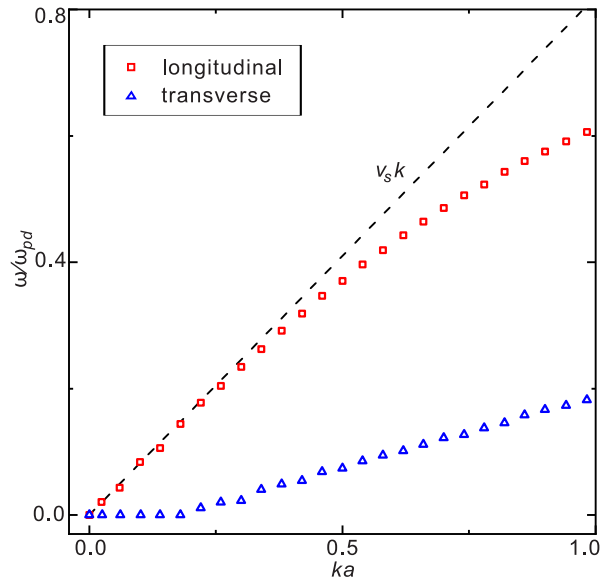}
\caption{\label{fig2} (Color online).
Longitudinal (a) and transverse (b) dispersion relations of the 2D dusty plasma liquid with $\kappa=1$ and ${\Gamma}/{\Gamma_m}=0.95$, as extracted from the data in Fig.~\ref{fig1}. The dashed line indicated the linear dispersion $\omega=v_s k$, where the value of the adiabatic sound speed $v_s$ is derived from the combination of the current conditions with the equation of state for 2D Yukawa liquids~\cite{Feng:2018}.}
\end{figure}

The longitudinal and transverse dispersion relations, determined using the DHO model fitting to the phonon spectra, are plotted in Fig.~\ref{fig2}. The slope of the longitudinal dispersion relation in Fig.~\ref{fig2} is approximately constant in the range of wavenumbers $ka \le 0.22$, just coinciding with the longitudinal adiabatic sound speed $v_s$ derived from hydrodynamics, as explained later. As the wavenumber further increases, $ka > 0.22$, the increase of the longitudinal frequency gradually slows down, and eventually the slope turns into negative after the peak frequency is reached, when $ka \approx 1.8$ from Fig.~\ref{fig1}. Unlike the longitudinal dispersion relation, however, the transverse dispersion relation increases approximately linearly in frequency when $ka \ge 0.18$. In addition, the frequency of the longitudinal dispersion relation is always greater than the transverse dispersion relation at the same wavenumber, as shown in Fig.~\ref{fig2}.

In Fig.~\ref{fig2}, we show the dispersion relations, $\omega_0(k)$ for both longitudinal and transverse modes. We draw a straight dashed line, whose slope just equals the corresponding adiabatic sound speed $v_s$ of the 2D dusty plasma liquid for $\kappa=1$ and ${\Gamma}/{\Gamma_m}=0.95$. The adiabatic sound speed is derived using the following expression from~\cite{Feng:2018,Li:2017}
\begin{equation}\label{cs}
{v_s}/{\sqrt{Q^2/2\pi\epsilon_0m a}}=\sqrt{-\frac{\kappa^4\gamma}{4}\left(\beta'/(\Gamma\kappa)+\alpha'\right)}.
\end{equation}
Substituting the conditions of $\kappa=1$, ${\Gamma}/{\Gamma_m}=0.95$ (${\Gamma}$=168.9), and using the results for the equation of state for 2D dusty plasma liquids in~\cite{Feng:2018,Li:2017}, we obtain the values for these coefficients $\alpha'=-2.67$, $\beta'=-2.64$, and the specific heat ratio $\gamma=1.004$, leading to the adiabatic sound speed of $v_s=0.82\sqrt{Q^2/2\pi\epsilon_0m a}$. The dispersion data are well fitted by the hydrodynamic dispersion $\omega=v_s k$ for small wavenumbers of $ka \le 0.22$. For larger $k$ values, the longitudinal dispersion relation never goes above the straight line $\omega=v_s k$, indicating that the so-called fast sound does not occur under the current conditions.

At the same time, the transverse dispersion relation in Fig.~\ref{fig2} exhibits the $k$-gap feature~\cite{Yang:2017, Baggioli:2020}, \textit{i.e.}, the real part of the frequency is zero below a certain wavenumber cutoff, which corresponds to roughly $ka \le 0.18$ in Fig.~\ref{fig2}. This indicates that transverse waves cannot propagate within this wavenumber range, which is what we do expect for liquids. Beyond a critical wavenumber, the transverse wave is able to propagate and the liquid behaves effectively like a solid. We will discuss the relation between fast-sound and the $k$-gap in more detail in the following sections.

\begin{figure}
\centering
\includegraphics{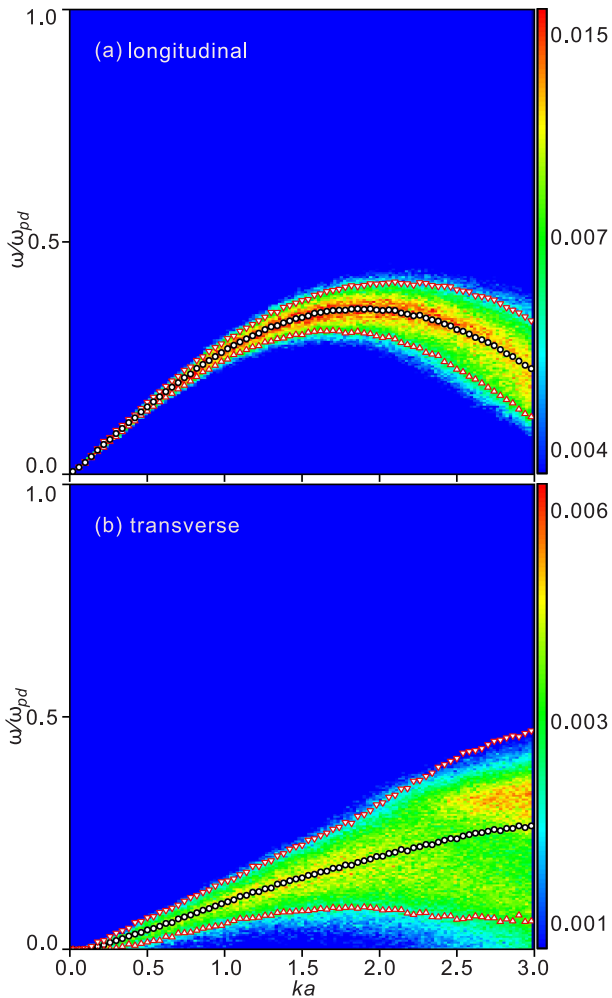}
\caption{\label{fig3} (Color online).
Longitudinal (a) and transverse (b) phonon spectra of the 2D dusty plasma liquid under $\kappa=2.5$ and ${\Gamma}/{\Gamma_m}=0.95$ calculated from equilibrium MD simulations.
}
\end{figure}

We also investigate the properties of the 2D dusty plasma liquid under different conditions, namely for $\kappa=2.5$ and ${\Gamma}/{\Gamma_m}=0.95$. The obtained phonon spectra is shown in Fig.~\ref{fig3}. As in the previous case, we perform the same averaging steps to improve the SNR of the phonon spectra. Furthermore, we also use the DHO model to determine the dispersion relations.

Due to the same value for the reduced coupling parameter~\cite{Huang:2023}, ${\Gamma}/{\Gamma_m}=0.95$, the phonon spectra in Fig.~\ref{fig3} are quite similar to those in Fig.~\ref{fig1}. When the wavenumber $ka \approx 1.8$, the longitudinal frequency reaches to its peak, nearly at the same wavenumber as in Fig.~\ref{fig1}. However, the peak frequency of the longitudinal spectra in Fig.~\ref{fig3} is clearly lower than that in Fig.~\ref{fig1}, due to the much larger value of $\kappa$. For larger $\kappa$ values, the 2D dusty plasma liquid is softer, reasonably leading to lower frequencies for both the longitudinal and transverse waves.

\begin{figure}
\centering
\includegraphics{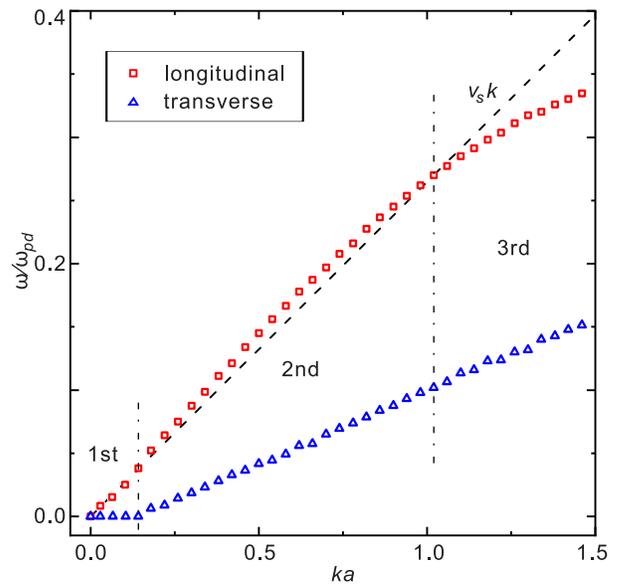}
\caption{\label{fig4} (Color online).
Longitudinal (a) and transverse (b) dispersion relations of a 2D dusty plasma liquid under the conditions of $\kappa=1$ and ${\Gamma}/{\Gamma_m}=0.95$ extracted from the data in Fig.~\ref{fig3}. The dashed line indicates the hydrodynamic dispersion $\omega=v_s k$, where $v_s$ is the adiabatic speed of sound. }
\end{figure}

In Fig.~\ref{fig4}, we plot the longitudinal and transverse dispersion relations, as determined by fitting the DHO model to the phonon spectra in Fig.~\ref{fig3}. On top of the numerical data, we also draw a dashed line with the slope of the corresponding adiabatic sound speed $v_s=0.26\sqrt{Q^2/2\pi\epsilon_0m a}$, derive from Eq.~(\ref{cs}) combined with the conditions of $\kappa=2.5$ and ${\Gamma}/{\Gamma_m}=0.95$.
 
As the main result of this paper, we discover the occurrence of fast sound in a 2D dusty plasma liquid from our simulations in Fig.~\ref{fig4}. While comparing the longitudinal dispersion relation with the straight line $\omega=v_s k$, we find that there are three distinctive ranges of wavenumber. In the first range, $ka\le 0.14$, the longitudinal dispersion relation nearly overlaps with $\omega=v_s k$, indicating that the longitudinal sound speed is the same as the adiabatic sound speed. In the second range, $0.14 < ka\le 1.02$, the frequencies corresponding to the longitudinal dispersion relation are higher than the hydrodynamic dispersion $\omega=v_s k$. In the last range, $ ka > 1.02$, the longitudinal dispersion bends down and the numerical data appear below the hydrodynamic dispersion $\omega=v_s k$. Importantly, in Fig.~\ref{fig4}, we observe that the onset of fast sound is concomitant with the $k$-gap in the transverse dispersion relation, around at $ka = 0.14$. We will return on this point in the next section.

\begin{figure}
\centering
\includegraphics{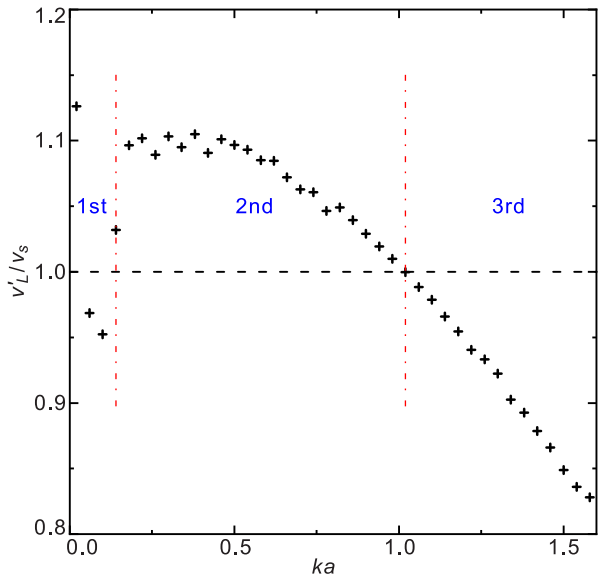}
\caption{\label{fig5} (Color online). Obtained longitudinal sound speed $v'_L=\omega/k$ from the dispersion relation data points in Fig.\ref{fig4}, normalized by the adiabatic sound speed $v_s$, for the 2D dusty plasma liquid for $\kappa=2.5$ and ${\Gamma}/{\Gamma_m}=0.95$. }
\end{figure}

To better quantify the fast sound, in Fig.\ref{fig5}, we plot the determined speed of longitudinal sound $v'_L=\omega/k$ ~\cite{Brazhkin:2013} normalized by the hydrodynamic adiabatic value $v_s$, as a function of the wavenumber $k$. We observe a maximal deviation of about $10\%$ with respect to the adiabatic sound speed $v_s$. Our obtained $10\%$ faster than $v_s$ longitudinal speed of sound is significant enough to draw the conclusion that the fast sound does occur under the corresponding conditions. The scattering of the simulation data for small values of $ka$ in Fig.\ref{fig5} is due to numerical inaccuracy. One obviously expected that $v_l'=v_s$ for $ka\ll1$. The appearance of the fast sound phenomenon will be confirmed later on using analytical approximate formulas for the elastic moduli of 2D dusty plasma liquids.

\section{Theoretical interpretation}\label{sec4}
In this section, we provide a simple theoretical interpretation of our numerical results using the Frenkel theory~\cite{Fomin:2016, Frenkel:1955}. In the hydrodynamic regime, the longitudinal sound speed in a liquid coincides with the adiabatic value, Eq.~\eqref{adi}, while transverse waves cannot be sustained since the static shear modulus vanishes. However, for smaller length scales, or equivalently for larger wavenumbers, typical liquids, including our studied 2D dusty plasma liquids, are able to sustain transverse waves, exhibiting solid-like properties \cite{Baggioli:2020}. Thus, while calculating the liquid longitudinal sound speed in this regime, the effect of the shear modulus should be included, leading to the expression of the longitudinal sound speed of Eq.~\eqref{lala}. At the same time, in this non-hydrodynamic regime, the transverse sound speed in liquids is expressed as $v_T^2 = \mu_\infty/\rho$. Combining the expressions for the longitudinal and transverse sound speeds, we obtain their ratio as
\begin{equation}\label{cl/cs}
\left(\frac{v'_L}{v_s}\right)^2=1+\frac{v_T^2}{v_s^2}.
\end{equation}

Using this theoretical framework, we confirm that our discovered fast sound is $\approx10\%$ beyond the adiabatic sound speed for the conditions show in Fig.~\ref{fig5}. Using Eq.~\eqref{cs}, combined with the conditions of $\kappa = 2.5$ and $\Gamma/\Gamma_m = 0.95$, we obtain the corresponding adiabatic sound speed $v_s=0.26\sqrt{Q^2/2\pi\epsilon_0m a}$~\cite{Feng:2018, Li:2017}. From~\cite{Wang:2019}, the instantaneous shear modulus $\mu_\infty$ of a 2D dusty plasma liquid is nearly the same as the shear modulus of the corresponding 2D dusty plasma solid with the same $\kappa$ value. As a result, we are able to directly use the transverse sound speed of a 2D dusty plasma crystal to derive the $\mu_\infty$ value of the corresponding 2D dusty plasma liquid. From the previous investigations~\cite{Wang:2019}, the transverse sound speed of 2D dusty plasma crystals $v_T$ can be phenomenologically fitted to
\begin{equation}\label{ct}
v_T/{\sqrt{Q^2/2\pi\epsilon_0m a}}=(0.374-0.0690\kappa^{1.11})/\sqrt{2}.
\end{equation}
Substituting $\kappa = 2.5$ into Eq.~\eqref{ct}, we obtain the corresponding transverse sound speed of $v_T = 0.13\sqrt{Q^2/2\pi\epsilon_0m a}$ for the dusty plasma crystal. At smaller length scales, the transverse sound speed of the corresponding dusty plasma liquid for $\kappa = 2.5$ should be approximately given by the same value. Thus, we derive the ratio of the longitudinal and transverse sound speeds in the 2D dusty plasma liquid of $\kappa = 2.5$ as ${v'_L}/{v_s}=1.11$. From Fig.~\ref{fig5}, we find that the maximum value of ${v'_L}/{v_s}$ is around $1.10$, well agreeing with our theoretical derivation here with only $\lesssim 1 \%$ difference.

\begin{figure}
\centering
\includegraphics{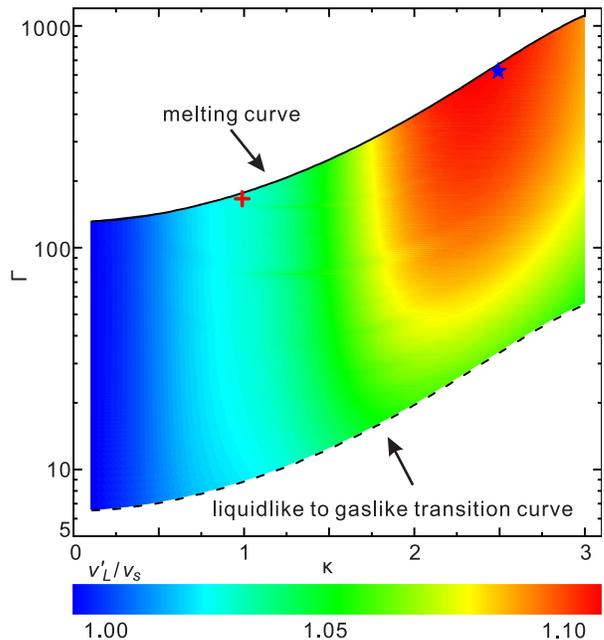}
\caption{\label{fig6} (Color online). Theoretical derived ratio of the longitudinal sound speed $v'_L$ to the adiabatic sound speed $v_s$ for 2D Yukawa liquids. The locations of the cross and the star symbols correspond to the parameters in Fig.~\ref{fig2} and  Fig.~\ref{fig4}, respectively.}
\end{figure}

To systematically study the fast sound in our studied system, in Fig.~\ref{fig6}, we plot our theoretical derived values of ${v'_L}/{v_s}$ for 2D dusty plasma liquids over a wide range of conditions. By combining Eqs.~(\ref{cs}), (\ref{cl/cs}), and (\ref{ct}) together, we derive the expression of ${v'_L}/{v_s}$ using $\kappa$ and ${\Gamma}$ as
\begin{equation}\label{cl/cskg}
{v'_L}/{v_s}=\sqrt{1-\frac{(0.374-0.0690\kappa^{1.11})^2}{\frac{\kappa^4\gamma}{2}\left(\beta'/(\Gamma\kappa)+\alpha'\right)}}.
\end{equation}
The solid curve on the top of Fig.~\ref{fig6} represents the melting conditions $\Gamma_m$ for 2D dusty plasmas, while the dashed curve on the bottom corresponding to the transition between the liquid-like to gas-like states of $\Gamma/\Gamma_m = 0.05$, also known as the Frenkel line~\cite{Huang:2023}. From~\cite{Baggioli:2020}, in the gas-like state for all supercritical fluids, like our studied dusty plasmas, transverse waves always cannot be sustained anymore, as a result, the fast sound cannot exist anymore~\cite{Fomin:2016}. However, in the liquid-like state, such as for the conditions between the solid and dashed curves in Fig.~\ref{fig6}, the fast sound may be observed, depending on its dependence on the system conditions, as we study next. 

From Fig.~\ref{fig6}, as $\kappa$ varies, ${v'_L}/{v_s}$ gradually increases until reaching its maximum when $\kappa \approx 2.4$ and then starts to decrease. From Fig.~\ref{fig6}, we also find that, when $\kappa \ge 1.8$, ${v'_L}/{v_s}$ is significantly dependent on the $\Gamma$ value, \emph{i.e.}, ${v'_L}/{v_s}$ increases noticeably with the $\Gamma$ value. Due to this dependence, the most significant deviation from the adiabatic speed should occur when $\kappa \approx 2.4$, especially at low temperatures near the melting curve, like for the conditions of Figs.~\ref{fig3} to~\ref{fig5}, as the star shown in Fig.~\ref{fig6}. However, from Fig.~\ref{fig6}, our predicted value of $v'_L/v_s$ is only $\approx 1.03$ for the conditions of Figs.~\ref{fig1} and~\ref{fig2}, probably too low to be easily identified from our simulations.

From our understanding, the variation trend of ${v'_L}/{v_s}$ in Fig.~\ref{fig6} can be interpreted by comparing the values for the adiabatic bulk modulus $K_s$ and the instantaneous shear modulus $\mu_\infty$ of our studied 2D dusty plasma liquids. As the screening parameter $\kappa$ increases, the shielding effect between particles significantly increases, so that the 2D dusty plasma liquid is much softer, leading to lower values for both $K_s$ and $\mu_\infty$. As $\kappa$ increases from $0.5$ to $3$, the instantaneous shear modulus for 2D dusty plasma liquids decreases by about one order of magnitude from Fig.~1 of~\cite{Wang:2019}. However, from Fig.~6 of~\cite{Feng:2018}, the adiabatic bulk modulus for 2D dust plasma liquids intensely decreases more than four orders of magnitude while $\kappa$ varies from $0.5$ to $3$. This significant difference in the decaying rates leads to an increasing importance of $\mu_\infty$ in the contribution to the longitudinal sound speed as $\kappa$ increases, resulting in the discovered larger values of $v'_L/v_s$ for higher $\kappa$ values. When the $\kappa$ value is too high with too strong screening effect, the 2D dusty plasma liquid is too close to the ideal gas, so that the transverse wave propagation is not so significant, leading to too insignificant feature of the fast sound to be detected.

\section{Conclusions}\label{sec5}
In conclusion, using MD simulations of 2D dusty plasmas in the liquid phase, we report the observation of the fast sound phenomenon in the dispersion of the longitudinal sound. Our results confirm that the fast sound phenomenon is directly linked to the emergence of solidity in the non-hydrodynamic regime of classical liquids. More precisely, we verify that the fast sound appears roughly at the same length scale of the $k-$gap phenomenon, which signals the onset of propagation for transverse shear waves. This strongly hints towards the conclusion that the two phenomena are just two manifestations of the same physics, as envisaged in some of the theoretical frameworks. In particular, we use the Frenkel theory to compare our numerical data with the corresponding theoretical predictions. These two are in agreement with only about $\lesssim 1 \%$ difference, confirming the validity of the theoretical framework even at the quantitative level. It is hard to believe this is just a mere coincidence, but it would be clearly helpful to pursue a larger exploration for different interaction potentials, dimensionality, etc to make a stronger case in this direction.

\subsection*{Aknowledgments}
The work was supported by the National Natural Science Foundation of China under Grant No. 12175159, the 1000 Youth Talents Plan, startup funds from Soochow University, and the Priority Academic Program Development of Jiangsu Higher Education Institutions. M.B. acknowledges the support of the Shanghai Municipal Science and Technology Major Project (Grant No.2019SHZDZX01) and the sponsorship from the Yangyang Development Fund.
 \bibliographystyle{apsrev4-1}

\begin{thebibliography}{16}
\bibitem{Landau:2013}
L.~Landau and E.~Lifshitz, {\it Fluid Mechanics: Course of Theoretical Physics} (Pergamon Press, Oxford, 1987), Vol.~6.

\bibitem{Boon:1991}
J.~P.~Boon and S.~Yip, {\it Molecular Hydrodynamics} (Dover Publications, New York, 1991).

\bibitem{Martin:1972}
P.~C.~Martin, O.~Parodi, and P.~S.~Pershan, Phys. Rev. A {\bf 6}, 2401~(1972).

\bibitem{Baggioli:2023}
M.~Baggioli and B.~Gout\'eraux, Rev. Mod. Phys. {\bf 95}, 011001~(2023).

\bibitem{Halperin:1969}
B.~I.~Halperin and P.~C.~Hohenberg, Phys. Rev. {\bf 188}, 898~(1969).

\bibitem{Chaikin:2000}
P.~M.~Chaikin and T.~C.~Lubensky, {\it Principles of Condensed Matter Physics} (Cambridge University Press, Cambridge, 2000).

\bibitem{Trachenko:2015}
K.~Trachenko and V.~V.~Brazhkin, Rep. Prog. Phys. {\bf 79}, 016502~(2015).

\bibitem{Baggioli:2020}
M.~Baggioli, M.~Vasin, V.~V.~Brazhkin, and K.~Trachenko, Phys. Rep. {\bf 865}, 1~(2020).

\bibitem{Yang:2017}
C.~Yang, M.~T.~Dove, V.~V.~Brazhkin, and K.~Trachenko, Phys. Rev. Lett. {\bf 118}, 215502~(2017).

\bibitem{Nosenko:2006}
V.~Nosenko, J.~Goree, and A.~Piel, Phys. Rev. Lett. {\bf 97}, 115001~(2006).

\bibitem{Noirez:2012}
N.~Laurence and B.~Patrick, Phys. Rev. Lett. {\bf 24}, 372101~(2012). 

\bibitem{Keyes:1997}
T.~Keyes, J. Phys. Chem. A {\bf 101}, 2921~(1997).

\bibitem{Stratt:1995}
R.~M.~Stratt, Acc. Chem. Res. {\bf 28}, 201~(1995).

\bibitem{Xiong:1999}
W.~X.~Li and T.~Keyes, J. Chem. Phys. {\bf 111}, 5503~(1995).

\bibitem{Zaccone:2021}
A.~Zaccone and M.~Baggioli, Proc. Natl. Acad. Sci. U.S.A. {\bf 118}, e2022303118~(2021).

\bibitem{Stamper:2022}
C.~Stamper, D.~Cortie, Z.~Yue, X.~Wang, and D.~Yu, J. Phys. Chem. Lett. {\bf 13}, 3105~(2022).

\bibitem{Madan:1993}
B.~Madan and T.~Keyes, J. Chem. Phys. {\bf 98}, 3342~(1993).

\bibitem{Baggioli:2021}
M.~Baggioli and A.~Zaccone, Phys. Rev. E {\bf 104}, 014103~(2021).

\bibitem{Hansen:2013}
J.~P.~Hansen and I.~R.~McDonald, {\it The Theory of Simple Liquids}, 2nd ed. (Elsevier Academic Press, Amsterdam, 1986).

\bibitem{Rahman:1974}
A.~Rahman and F.~H.~Stillinger, Phys. Rev. A {\bf 10}, 368~(1974).

\bibitem{Teixeira:1985}
J.~Teixeira, M.~C.~Bellissent-Funel, S.~H.~Chen, and B.~Dorner, Phys. Rev. Lett. {\bf 54}, 2681~(1985).

\bibitem{Sette:1995}
F.~Sette, G.~Ruocco, M.~Krisch, U.~Bergmann, C.~Masciovecchio, V.~Mazzacurati, G.~Signorelli, and R.~Verbeni, Phys. Rev. Lett. {\bf 75}, 850~ (1995).

\bibitem{Ruocco:2008}
G.~Ruocco and F.~Sette, Condens. Matter Phys. {\bf 11}, 29~(2008).

\bibitem{Balucani:1993}
U.~Balucani, G.~Ruocco, A.~Torcini, and R.~Vallauri, Phys. Rev. E {\bf 47}, 1677~(1993).

\bibitem{Cunsolo:1999}
A.~Cunsolo, G.~Ruocco, F.~Sette, C.~Masciovecchio, A.~Mermet, G.~Monaco, M.~Sampoli, and R.~Verbeni, Phys. Rev. Lett. {\bf 82}, 775~(1999).

\bibitem{Cunsolo:2015}
A.~Cunsolo, Adv. Cond. Matter Phys. {\bf 2015}, 137435~(2015). 

\bibitem{Bryk:2005}
T.~Bryk and I.~Mryglod, Phys. Rev. B {\bf 71}, 132202~(2005).

\bibitem{Bosse:1986}
J.~Bosse, G.~Jacucci, M.~Ronchetti, and W.~Schirmacher, Phys. Rev. Lett. {\bf 57}, 3277~(1986).

\bibitem{Demmel:2004}
F.~Demmel, S.~Hosokawa, M.~Lorenzen, and W.~C.~Pilgrim, Phys. Rev. B {\bf 69}, 012203~(2004).

\bibitem{Alvarez:1998}
M.~Alvarez, F.~J.~Bermejo, P.~Verkerk, and B.~Roessli, Phys. Rev. Lett. {\bf 80}, 2141~(1998).

\bibitem{Campa:1988}
A.~Campa and E.~G.~D.~Cohen, Phys. Rev. Lett. {\bf 61}, 853~(1988).

\bibitem{Ruzicka:2004}
B.~Ruzicka, T.~Scopigno, S.~Caponi, A.~Fontana, O.~Pilla, P.~Giura, G.~Monaco, E.~Pontecorvo, G.~Ruocco, and F.~Sette, Phys. Rev. B {\bf 69}, 100201~(2004).

\bibitem{Bencivenga:2014}
F.~Bencivenga and D.~Antonangeli, Phys. Rev. B {\bf 90}, 134310~(2014).

\bibitem{Sabirov:2010}
L.~M.~Sabirov, D.~I.~Semenov, and K.~S.~Khaidarov, Phys. Wave Phenomena, {\bf 18}, 159~2010.

\bibitem{Bryk:2010}
T.~Bryk, I.~Mryglod, T.~Scopigno, G.~Ruocco, F.~Gorelli, and M.~Santoro, J. Chem. Phys. {\bf 133}, 024502~(2010).

\bibitem{Santucci:2006}
 S.~C.~Santucci, D.~Fioretto, L.~Comez, A.~Gessini, and C.~Masciovecchio, Phys. Rev. Lett. {\bf 97}, 225701~(2006).
 
\bibitem{Bryk:2014}
T.~Bryk, F.~Gorelli, G.~Ruocco, M.~Santoro, and T.~Scopigno, Phys. Rev. E {\bf 90}, 042301~(2014).

\bibitem{Bryk:1997}
T.~Bryk, I.~Mryglod, and G.~Kahl, Phys. Rev. E {\bf 56}, 2903~(1997).

\bibitem{Bryk:2002}
T.~Bryk and I.~Mryglod, J. Phys. Condens. Matter {\bf 14}, L445~(2002).

\bibitem{Bryk:2011}
T.~Bryk, Eur. Phys. J. Spec. Top. {\bf 196}, 65~(2011).

\bibitem{Cohen:1971}
C.~Cohen, J.~W.~H.~Sutherland, and J.~M.~Deutch, Phys. Chem. Liquids {\bf 2}, 213~(1971).

\bibitem{Schepper:1988}
I.~M.~de Schepper, E.~G.~D.~Cohen, C.~Bruin, J.~C.~van Rijs, W.~Montfrooij, and L.~A.~de Graaf, Phys. Rev. A {\bf 38}, 271~(1988).

\bibitem{Mryglod:1999}
I.~Mryglod, J. Phys. Stud. {\bf 3} 33~(1999). 

\bibitem{Baggioli:2022}
M.~Baggioli, M.~Landry, and A.~Zaccone, Phys. Rev. E {\bf 105}, 024602~(2022).

\bibitem{Frenkel:1955}
J.~Frenkel, {\it Kinetic Theory of Liquids}, (Oxford University Press, New York, 1946).

\bibitem{Thomas:1996}
H.~M.~Thomas and G.~E.~Morfill, Nature {\bf 379}, 806~(1996).

\bibitem{Lin:1996}
L.~I, W.~T.~Juan, C.~H.~Chiang, and J.~H.~Chu, Science {\bf 272}, 1626~(1996).

\bibitem{Melzer:1996}
A.~Melzer, A.~Homann, and A.~Piel, Phys. Rev. E {\bf 53}, 2757~(1996).

\bibitem{Merlino:2004}
R.~L.~Merlino and J.~A.~Goree, Phys. Today {\bf 27}, 32~(2004).

\bibitem{Kalman:2004}
G.~J.~Kalman, P.~Hartmann, Z.~Donk\'o, and M.~Rosenberg, Phys. Rev. Lett. {\bf 92}, 065001~(2004).

\bibitem{Morfill:2009}
G.~E.~Morfill and A.~V.~Ivlev, Rev. Mod. Phys. {\bf 81}, 1353~(2009).

\bibitem{Bonitz:2010}
M.~Bonitz, C.~Henning, and D.~Block, Rep. Prog. Phys. {\bf 73}, 066501~(2010).

\bibitem{Thomas:2005}
E.~Thomas, Jr., and J.~Williams, Phys. Rev. Lett. {\bf 95}, 055001~(2005).

\bibitem{Feng:2010}
Y.~Feng, J.~Goree, and B.~Liu, Phys. Rev. Lett. {\bf 104}, 165003~(2010).

\bibitem{Feng:2012}
Y.~Feng, J.~Goree, and B.~Liu, Phys. Rev. Lett. {\bf 109}, 185002~(2012).

\bibitem{Konopka:2000}
U.~Konopka, G.~E.~Morfill, and L.~Ratke, Phys. Rev. Lett. {\bf 84}, 891~(2000).

\bibitem{Nunomura:2002}
S.~Nunomura, J.~Goree, S.~Hu, X.~Wang, A.~Bhattacharjee, and K.~Avinash, Phys. Rev. Lett. {\bf 89}, 035001~(2002).

\bibitem{Nosenko:2008}
V.~Nosenko, S.~Zhdanov, A.~Ivlev, G.~Morfill, J.~Goree, and A.~Piel, Phys. Rev. Lett. {\bf 100}, 025003~(2008).

\bibitem{Hartmann:2010}
P.~Hartmann, A.~Douglass, J.~C.~Reyes, L.~S.~Matthews, T.~W.~Hyde, A.~Kov\'acs, and Z.~Donk\'o, Phys. Rev. Lett. {\bf 105}, 115004~(2010).

\bibitem{Couedel:2010}
L.~Cou\"edel, V.~Nosenko, A.~V.~Ivlev, S.~K.~Zhdanov, H.~M.~Thomas, and G.~E.~Morfill, Phys. Rev. Lett. {\bf 104}, 195001~(2010).

\bibitem{Kahler:2012}
H.~K\"ahlert, J.~Carstensen, M.~Bonitz, H.~L\"owen, F.~Greiner, and A.~Piel, Phys. Rev. Lett. {\bf 109}, 155003~(2012).
 
\bibitem{Hartmann:2014}
P.~Hartmann, A.~Z.~Kov\'acs, A.~M.~Douglass, J.~C.~Reyes, L.~S.~Matthews, and T.~W.~Hyde, Phys. Rev. Lett. {\bf 113}, 025002~(2014).

\bibitem{Gogia:2017}
G.~Gogia and J.~C.~Burton, Phys. Rev. Lett. {\bf 119}, 178004~(2017).

\bibitem{Hartmann:2019}
P.~Hartmann, J.~C.~Reyes, E.~G.~Kostadinova, L.~S.~Matthews, T.~W.~Hyde, R.~U.~Masheyeva, K.~N.~Dzhumagulova, T.~S.~Ramazanov, T.~Ott, H.~K\"ahlert, M.~Bonitz, I.~Korolov, and Z.~Donk\'o, Phys. Rev. E {\bf 99}, 013203~(2019).

\bibitem{ChengRan:2019}
C.-R.~Du, V.~Nosenko, H.~M.~Thomas, Y.~F.~Lin, G.~E.~Morfill, and A.~V.~Ivlev, Phys. Rev. Lett. {\bf 123}, 185002~(2019).

\bibitem{Nunomura:2003}
S.~Nunomura, S.~Zhdanov, G.~Morfill, and J.~Goree, Phys. Rev. E {\bf 68}, 026407~(2003).

\bibitem{Goree:2012}
J.~Goree, Z.~Donk\'o, and P.~Hartmann, Phys. Rev. E {\bf 85}, 066401~(2012).

\bibitem{Piel:2006}
A.~Piel, V.~Nosenko, and J.~Goree, Phys. Plasmas {\bf 13}, 042104~(2006).

\bibitem{Nunomura:2000}
S.~Nunomura, D.~Samsonov, and J.~Goree, Phys. Rev. Lett. {\bf 84}, 5141~(2000).

\bibitem{Nunomura:2002_2}
S.~Nunomura, J.~Goree, S.~Hu, X.~Wang, A.~Bhattacharjee, and K.~Avinash, Phys. Rev. E {\bf 65} 066402~(2002).

\bibitem{Nunomura:2005}
S.~Nunomura, D.~Samsonov, S.~Zhdanov, and G.~Morfill, Phys. Rev. Lett. {\bf 95}, 025003~(2005).

\bibitem{Feng:2012_2}
Y.~Feng, J.~Goree, and B.~Liu, Phys. Rev. E {\bf 86}, 056403~(2012).

\bibitem{Juan:1998}
W.-T.~Juan and I.~Lin, Phys. Rev. Lett. {\bf 80}, 3073~(1998).

\bibitem{Feng:2008}
Y.~Feng, J.~Goree, and B.~Liu, Phys. Rev. Lett. {\bf 100}, 205007~(2008).

\bibitem{Ohta:2000}
H.~Ohta and S.~Hamaguchi, Phys. Rev. Lett. {\bf 84}, 6026~(2000).

\bibitem{Liu:2003}
Y.~Liu, B.~Liu, Y.~Chen, S.-Z.~Yang, L.~Wang, and X.~Wang, Phys. Rev. E {\bf 67}, 066408~(2003).

\bibitem{Wang:2001}
X.~Wang, A.~Bhattacharjee, and S.~Hu, Phys. Rev. Lett. {\bf 86}, 2569~(2001).

\bibitem{Samsonov:2004}
D.~Samsonov, S.~Zhdanov, R.~Quinn, S.~Popel, and G.~Morfill, Phys. Rev. Lett. {\bf 92} ,255004~(2004).

\bibitem{Kananovich:2020}
A. Kananovich and J. Goree, Phys. Rev. E {\bf 101}, 043211~(2020).

\bibitem{Fortov:2005}
V.~Fortov, A.~Ivlev, S.~Khrapak, A.~Khrapak, and G.~Morfill, Phys. Rep. {\bf 421}, (2005)~1.

\bibitem{Ohta:2000_2}
H.~Ohta and S.~Hamaguchi, Phys. Plasmas {\bf 7}, 4506~(2000).

\bibitem{Sanbonmatsu:2001}
K.~Sanbonmatsu and M.~Murillo, Phys. Rev. Lett. {\bf 86}, 1215~(2001).

\bibitem{Hartmann:2005}
P.~Hartmann, G.~J.~Kalman, Z.~Donk\'o, and K.~Kutasi, Phys. Rev. E {\bf 72}, 026409~(2005).

\bibitem{Huang:2023}
D.~Huang, M.~Baggioli, S.~Lu, Z.~Ma, and Y.~Feng, Phys. Rev. Res. {\bf 5}, 013149~(2023).

\bibitem{Liu:2003_2}
B.~Liu, K.~Avinash, and J.~Goree, Phys. Rev. Lett. {\bf 91}, 255003~(2003).

\bibitem{Feng:2016}
Y.~Feng, W.~Lin, W.~Li, and Q.~Wang, Phys. Plasmas {\bf 23}, 093705 (2016) {\bf 23}, 119904~(2016).

\bibitem{Li:2018}
W.~Li, D.~Huang, K.~Wang, C.~Reichhardt, C.~J.~O.~Reichhardt, M.~Murillo, and Y.~Feng, Phys. Rev. E {\bf 98}, 063203~(2018).

\bibitem{Yurchenko:2018}
S.~O.~Yurchenko, K.~A.~Komarov, N.~P.~Kryuchkov, K.~I.~Zaytsev, and V.~V.~Brazhkin, J. Chem. Phys. {\bf 148}, 134508~(2018).

\bibitem{Khrapak:2018}
S.~A.~Khrapak, N.~P.~Kryuchkov, L.~A.~Mistryukova, A.~G.~Khrapak, and S.~O.~Yurchenko, J. Chem. Phys. {\bf 149}, 134114~(2018).


\bibitem{Khrapak:2019}
S.~A.~Khrapak, A.~G.~Khrapak, N.~P.~Kryuchkov, and S.~O.~Yurchenko, J. Chem. Phys. {\bf 150}, 104503~(2019).

\bibitem{Feng:2018}
Y.~Feng, D.~Huang, and W.~Li, Phys. Plasmas {\bf 25}, 057301~(2018).

\bibitem{Li:2017}
W.~Li, W.~Lin, and Y.~Feng, Phys. Plasmas {\bf 24}, 043702~(2017).

\bibitem{Brazhkin:2013}
V.~V.~Brazhkin, Yu.~D.~Fomin, A.~G.~Lyapin, V.~N.~Ryzhov, E.~N.~Tsiok, and K.~Trachenko, Phys. Rev. Lett. {\bf 111}, 145901~(2013).

\bibitem{Fomin:2016}
Y.~D.~Fomin, V.~N.~Ryzhov, E.~N.~Tsiok, V.~V.~Brazhkin, and K.~Trachenko, J. Phys.: Condens. Matter {\bf 28}, 43LT01~(2016).

\bibitem{Wang:2019}
K.~Wang, D.~Huang, and Y.~Feng, Phys. Rev. E {\bf 99}, 063206~(2019).

\end{thebibliography}

\end{document}